\documentstyle[12pt]{article}
\begin{document}
\vspace{20mm}

\title{Multifractal Interpolation of Universal Multifractals}
\author{\sl V.G.Bar'yahtar, V.Yu.Gonchar, D.Schertzer*, V.V.Yanovsky \\
  \\
  \it Laboratory for Turbulence Research  \\
\it Institute for Single Crystals National Acad.Sci.Ukraine,\\
\it Lenin ave. 60, Kharkov 310001,Ukraine,\\
\it {*}Laboratoire de Modelisation en Mecanique,\\
 \it place Jussieu, F-75252,Paris, France }
  \date{}
 \maketitle

\begin{abstract}
Basing on invariant properties of universal multifractals we propose a
simple algorithm for interpolation of multifractal densities. This
algorithm admits generalizations to a multidimensional case. Analytically
obtained are multifractal characteristic of the function interpolating
initial data. We establish the relation between the parameter existing in
algorithm and the Levy index $\alpha$ which is the main index for scaling
function of universal multifractals
\end{abstract}
 \section{Introduction} Construction of the models for
complex highly variable natural processes is closely associated with the
study of their multifractal characteristics. Popular methods of averaging
allows one to study smoothed characteristics only and lead to the loss of
information about variability of the processes on scales less than the
scales of averaging. Therefore, the problem is how to construct the
function averaged over the scales $\epsilon << \epsilon_0$ in the limit
$\epsilon \rightarrow 0$ with the help of smoothed experimental data (that
is, averaged over some given scale $\epsilon_0$ ).  In fact, it implies the
reconstruction of variability at small scales.  Obviously, this problem
has no unique solution without additional requirements and principles of
such an interpolation. Examples of fractal interpolations with different
principles are presented, for example, in Refs. \cite{bar},
\cite{the}.  In the paper we use other principles which are related
to the possibility to return from interpolating function to the smoothed
one by averaging over scales. At the same time, returning from an
arbitrary interpolation step (that is, from the interpolating function on
an arbitrarily small scale) we should get an initial function. The second
principle is associated with the idea of similarity in a broad sense, and
implies the conservation of memory about small-scale variability of
unknown strict function in smoothed data.  Similar principles are already
widely used in various cascade models of turbulence (see,
e.g.$\alpha$-model \cite{fri}, \cite{pal}, and $\beta$-model
\cite{nov}, \cite{man} ).  Besides, we use a natural requirement
that interpolated (smoothed) function possesses definite multifractal
characteristics. This requirement is general for all methods of fractal
 \cite{bar}, \cite{the} and multifractal interpolation.  We use
 the momentum scaling function $K(q)$ as a multifractal characteristic. It
 is related by the known way with the spectrum of singularity $f( \alpha
 )$ \cite{mcc}, \cite{gra}, \cite{mct} and the mass power \cite{mea},
\cite{met}.  At least, we require the simplicity of the interpolation
procedure, this requirement being hardly defined.  We propose a simple and
fast interpolation algorithm which allows one to make its generalization
on a multidimensional case.  The multifractal characteristics of the
function interpolating initial data are obtained analytically. It is
shown, that multifractals arising as a result of interpolation, belongs to
universal multifractals  \cite{sch} associated with the deep properties of
stochastic processes. The proposed interpolation method is based on the
existence of the invariants (that is, the characteristics independent of
scales) obtained in the paper for the universal multifractals.

\section{Formulation of the problem and the algorithm}
Let us consider the process of smoothing the function $\rho (x)$. We
assume without loss of generality that the unit interval [0,1] is the
definition region of this function. Then, $\rho (x)$ is assumed as
integrable.  We divide the definition region in to N equal segments
$[0,1] = \bigcup_{n=1}^{N}I_n $ , $I_n = ((n - 1) \epsilon ,
n \epsilon )$ where $\epsilon = 1/N $.  We defined the step-like function
$\rho_\epsilon (x)$ which has a non-zero constant value at each interval
$I_n $,
   \begin{eqnarray}
 \rho_\epsilon (x) =
 \frac{\int\limits_{n \epsilon}^{(n+1)\epsilon}\rho (x)dx}{\epsilon} ,\,\,
 x\in I_n \end{eqnarray}
  Where $x\in[0,1]$ .
 The function $\rho_\epsilon (x)$ smoothes the function $\rho (x)$ at scale
 $\epsilon $ .  Successive increasing the scales of smoothing allows us
  pass to more and more smoothed description of $\rho(x)$.  The function
 $\rho_1 (x)$ (L=1) is the last in the hierarchy of smoothed descriptions,
 \begin{eqnarray} \rho_1 (x) = \frac{\int\limits_{0}^{L}\rho (x)dx}{L}
 \equiv \langle \rho \rangle \end{eqnarray}
  where $\langle ...\rangle$ the parentheses
imply spatial averaging. The function $\rho_\epsilon$, $x\in [0,1]$ form
a completely ordered set.  It is easily proved that $\langle ...\rangle$
 is invariant at all stages of smoothing. Indeed,

 \begin{eqnarray} \langle \rho_\epsilon (x) \rangle =
 \frac{\int\limits_{0}^{L} \rho_\epsilon (x)dx}{L} =
 \frac{\sum\limits_{i=0}^{N-1} \int\limits_{i\epsilon}^{(i+1)\epsilon}
  \rho (x)dx \epsilon}{L \epsilon} = \nonumber \\
  = \frac{1}{L}
  \sum\limits_{i=0}^{N-1} \int\limits_{i\epsilon}^{(i+1)\epsilon}\rho
  (x)dx = \frac{1}{L} \int \limits_{0}^{L} \rho (x)dx = \langle \rho
  \rangle
   \end{eqnarray}

Therefore, the mean value of any smoothed function $\rho_\epsilon (x)$
coincides with that of the initial function.  The described process of
smoothing simulated the main features of experimental data treatment. Let
us formulate the problem. We assume that the initial density $\rho (x)$ is
a multifractal with given momentum scaling function $ K(q) $. Knowing
smoothed description $\rho_{\epsilon_0} (x)$ , we try to restore
$\rho_\epsilon (x)$ at $\epsilon << \epsilon_0 $ with the same momentum
function.  In other words, we create a model of the function$\rho (x)$ with
conserved properties of variability at small scales $\epsilon \rightarrow
0$.  The model $\tilde{\rho} (x) = \lim\limits_{\epsilon\to 0}\rho_\epsilon
 (x)$ is named as interpolating function of the smoothed density
$\rho_{\epsilon_0}$ .  According to the meaning of smoothing, it is not
necessary for $\tilde{\rho} (x)$ to get the fixed values which coincide
with the corresponding values $\rho_{\epsilon_0} (x) $ of at the definite
$x$.  We expect that this problem has no unique solution, because the
integration procedure in Eq.(1) implies the loss of information about the
variability of $\rho (x)$ on scales less than $\epsilon$ .  Besides, after
the reconstruction of its variability, the interchanges of particular
segments of the function are assumed, those do not influence their
multifractal characteristics. Therefore, let us formulate the particular
algorithm of interpolation and discuss its properties.
 We assume that
given smoothed step-like function $\rho_{\frac{1}{p}} (x)$ (p is some
integer) has some value not equal zero at each interval $\frac{1}{p}$
long.  The absence of zero values is not principal and is assumed for the
simplicity of subsequent analysis only. Therefore, we know
$\rho_\epsilon (x)$ at $\epsilon > \frac{1}{p}$, and, in particular,
$\rho_1 = \langle \rho \rangle$ .  Below we assume  $\langle \rho \rangle$
without loss of generality.
  We describe the algorithm as successive
iterations of $\rho_{\frac{1}{p}} (x)$ , each of them consisting in 3
simple operations:

 1.  The contraction of from the interval [0,1] onto
the intervals  [0,$ \frac{1}{p^2}$],
 [$\frac{1}{p^2}$ ,$ \frac{2}{p^2}$],
[$\frac{p^2 - 1}{p^2}$ ,1]  and the construction of the
function with the period $\frac{1}{p^2}$. The periodic function repeats on
its period the behavior of $\rho_{\frac{1}{p}} (x)$ at the whole interval
[0,1], that is it coincides with $\rho_{\frac{1}{p}} (x p^2)$ (operation
{\bf P}).

 2.  The periodic function obtained at the 1-st step is raised to the
 $\nu$ -th power (operation {\bf R}). In general case the power $\nu$  can
 depend on the number n of the interval $I_n =
 [\frac{n-1}{p^2} , \frac{n}{p^2} ] $ , and on the iteration step.  We
 restrict ourselves to the case $\nu$=const, for simplicity.

  3. The mean value
 is normalized (operation {\bf N}), that is, the mean value at each
 interval $I_n$ is fitted to the corresponding value of $\rho_\frac{1}{p}
 (x)$ at the same interval.  This is done by multiplication of the values
 at the interval $I_n$ on the value of $\rho_{\frac{1}{p}} (x)$ at $x\in
 {I_n}$, and subsequent division of the function, constructed after the
 steps 1 and 2, on the mean value at this interval.

  A single iteration
 step of the function $\rho_{\frac{1}{p}} (x)$  consists in applying
  operation {\bf NRP} to it. An
 analogous transition is realized from the iteration step at the scale of
 smoothing $\epsilon_k$ to k+1 - st step at the scale
 $\epsilon_{(k+1)}$, that is,

 \begin{eqnarray} \rho_{\epsilon_{k+1}} (x) = {\bf NRP} \rho_{\epsilon_k}
 (x) \end{eqnarray}

It is seen, that during such an algorithm $\epsilon_{k+1} = \epsilon^3_k$
while the number of intervals $N_{k+1} = \frac{1}{\epsilon_{k+1}}$ is
increased according to $N_{k+1} = N^3_k$.  It implies that, if we have p
intervals initially, then on the k-th step $N_k = p^{3^k}$ , that is, the
variability of the function $\rho_{\epsilon_k} (x)$  grows hyper
exponentially.  The operations {\bf P,R} are based on scale similarity
properties of $\rho (x)$ and conservation of "memory" during the
smoothing.  Besides, {\bf R} has a control parameter $\nu$ influencing the
multifractal properties of the interpolating function. The operation {\bf
N} ensures conservation of the mean values and is related to the invariant
of smoothing described above.  The formulated algorithm is exclusively
simple, easily realizable on computers and allows one to provide a
complete analytical analysis of the interpolating functions.

\section{Analysis of multifractal properties of interpolating functions}
Let us proceed with the study of the multifractal characteristics of
 the function $\rho_{\epsilon} (x)$, which are constructed according to
 the algorithm presented above. It is convenient to pass from
 $\rho_{\epsilon_0} (x)$ to the functions defined on integers. We give
 a natural ordinal number n to the segments of the unit segment
 subdivision with length $\epsilon_k$, and associate the value of a
 step-like function $\rho_{\epsilon_k} (x)$ to the function $\rho_k (n)$
 at this segment, that is,
 \begin{eqnarray} \rho_k (n) = \rho_{\epsilon_k} (x \mid x\in
 [(n-1)\epsilon_k , n \epsilon_k ] ) \end{eqnarray}

Then, the index k implies the number of iteration, while n=1,2,...,$N_k$.
Here $N_k$ is a number of the unit segment subdivision intervals with the
length $\epsilon_k$ at the k-th iteration step.  We formulate in an
explicit form the rule of the transition ($\bf NPR$) from $\rho_k (m)$
to $\rho_{k+1} (n)$ (m=1,2,...,$N_k$; n=1,2,...,$N_{k+1}$).  It is easily
proved that
\begin{eqnarray}\rho_{k+1} (n) = \frac{\rho_k{}^\nu (n \bmod{N_k}
)}{M_\nu (k)} \rho_k (1 + \Bigl[ \frac{n}{N_k{}^2} \Bigr] )
\end{eqnarray} Here the standard notations are used; the square brackets
denote an integer part of a number, (j + p $N_k$ ) mod $N_k$ =j if, $0\le
j \le N_k$ while p is an integer;n=1,..,$N_k$. $M_\nu (k)$  is a $\nu$ -th
moment of $\rho_{\epsilon_k} (x)$, that is,

 \begin{eqnarray} M_{\nu} (k) =\frac{\int_{0}^{1} \rho_{\epsilon_k}
 (x)dx}{1} = \epsilon_k \sum\limits_{n-1}^{N_k} \rho_k{}^\nu (n) =
 \frac{1}{N_k} \sum\limits_{n=1}^{N_k} \rho_k{}^\nu (n)
 \end{eqnarray}

Eq. (6) allows us to calculate easily arbitrary moments of the
introduced density $\rho_k (n)$. As an example, we prove the conservation
of the mean density on an arbitrary iteration step. By definition, the
  mean density at (k+1)-st iteration step is
 $$ M_1 (k+1) = < \rho_{k+1} > = \frac{1}{N_{k+1}}
   \sum\limits_{n=1}^{N_{k+1}} \rho_{k+1} (n) $$

Inserting the expression for $\rho_{k+1} (n)$ see Eq. (6), and taking
into account that $N_{k+1} =N_k^3$, after partial summation we get

$$<\rho_{k+1}> = \frac{1}{N_k{}^3} N_k \frac{\sum\limits_{n=1}^{N_k}
\rho_k{}^\nu (n)}{M_\nu (k)} \sum\limits_{n=1}^{N_k} \rho_k (n) =
\frac{M_\nu (k)}{M_\nu (k)} \frac{1}{N_k} \sum\limits_{n=1}^{N_k} \rho_k
(n) \equiv <\rho_k >$$

Therefore, when using the described algorithm, the mean value is
 conserved at arbitrary iteration step. We assume < $\rho$> = 1 without
 loss of generality.
Then, we determine how multifractal characteristics of
 the density $\rho_k (n)$ (that is, those of $\rho_{\epsilon_k} (x)$)
 depend on the free parameter $\nu$. For this purpose it is necessary to
 calculate more general fractional order moments of the constructed
   function $\rho_k (n)$.  At the first step, using Eq.(6), we establish
   the relation between the q-th moment at the (k+1)-st iteration step
    with that at the k-th iteration step.  By definition

    \begin{eqnarray}M_q (k+1) =\frac{1}{N_{k+1}}
    \sum\limits_{n=1}^{N_{k+1}} \rho_{k+1}{}^q (n) \end{eqnarray}

Using Eq. (6) and $N_{k+1} = N_k^3$, after partial summation we get

    \begin{eqnarray}M_q (k+1) =\frac{M_{q\nu} (k)}{M_\nu{}^q (k)} M_q (k)
    \end{eqnarray}

Here the standard notations of the moments are used, see Eqs. (7),(8).
 We use the momentum formalism \cite{sch}, \cite{sct} in order
 to determine multifractal characteristics of the interpolating function.
 The main definition of this formalism is that of the momentum scaling
 function $K(q)$ of multifractal density $\rho_\epsilon (x)$ at scale
  resolution $\epsilon$:  \begin{eqnarray}< \rho_\epsilon{}^q (x) > =
  \epsilon^{-K(q)} \end{eqnarray}

The function $K(q)$ is obeyed the conditions $K(0)=K(1)=0$ which are the
consequences of the definition (10) and the relation < $\rho$ >  = 1. It is
 worthwhile to note that $K(q)$ is independent of scales , and related
 by the known way to the mass power $\tau (q)$ \cite{mat}:
$$\tau (q) = 1 - q + K(q)$$
The mass power is related by a Legendre transform to the singularity
 function $f(\alpha)$, see for example, Ref \cite{mat}. Returning
 to Eq.(9), we consider it as a functional equation for the moments of
  interpolating function. We look for the solution in the form
\begin{eqnarray}M_q (k) = \epsilon_k^{-K(q)}
\end{eqnarray}

Then, for the scaling function K(q) we get the equation
$$ \epsilon_{k+1}^{-K(q)} = \epsilon_k^{qK(\nu ) - K(q\nu ) - K(q)}$$

Using $\epsilon_{k+1} = \epsilon_k^3$, it is easy to exclude the
dependence on scales and get the functional equation for $K(q)$;

\begin{eqnarray}2 K(q) = K(q \nu ) - qK(\nu )
\end{eqnarray}
 The absence of scale dependence in Eq.(13) implies that we really get
 multifractal density $\rho (x)$ as a result of interpolation. To find the
solution of Eq. (13) we make a substation $K(q) = \psi (q)$ and get the
equation for $\psi (q)$:
$$2\psi (q) = \nu ( \psi (q \nu ) - \psi ( \nu ) )$$

We look for the solution in the class of differentiated functions. We
 differentiate this equation over q and get the following equation for
  the derivative $\psi{}{}^{\prime} (q)$:
$$2\psi{}{}^{\prime} (q) = \nu^2 \psi{}{}^{\prime} (q\nu )$$

The solution at $ q > 0 $ is
$$\psi{}{}^{\prime} (q) = K_0 q^{\frac{\ln 2}{\ln \nu} - 2}$$

Integrating this equation and returning to $ K(q) $ we get ultimately

\[ K(q) = \left\{\begin{array}{rl}
\frac{K_0}{\frac{\ln 2}{\ln \nu} - 1} q^{\frac{\ln 2}{\ln \nu}} + c_1
q, & \mbox{if} \, \, \nu \neq 2\\ qK_0 \ln q + c_1 q, &\mbox{if} \, \, \nu
= 2 \end{array} \right. \]

Using natural boundary conditions $K(0)=K(1)=0$, we exclude an arbitrary
constant $c_1$:
\begin{eqnarray} K(q) = \left\{\begin{array}{rl}
\frac{K_0}{\alpha - 1} ( q^\alpha - q )
, & \mbox{if} \, \, \nu \neq 2\\ K_0 q \ln q , &\mbox{if} \, \,  \nu = 2
\end{array} \right.
\end{eqnarray}

where $\alpha =\frac{\ln 2}{\ln \nu}$, $K_0$ is an arbitrary constant.
It is worthwhile to note that Eq. (13) coincides with the known $K (q)$
for the so-called universal multifractals \cite{sch}, \cite{sct}.
Extraction of this class has a deep mathematical origin associated with
the existence of limit theorems for the distributions of the sums of
independent stochastic quantities \cite{lev}. From the other hand,
just this class of multifractals is the main pretender to the description
of real natural processes and objects. This fact is supported by
  experimental investigations of various natural phenomena \cite{sch},
   \cite{mat}.

     Therefore, we have shown that
interpolating function $\rho_{\epsilon} (x)$ is universal multifractal,
the main characteristic $\alpha = \frac{\ln 2}{\ln \alpha}$ of it being
the known function of the control parameter $\nu$. Varying $\nu$, it is
possible to construct various interpolations of experimental data with
various multifractal characteristics.  It is worthwhile to note that this
analysis, actually points to the existence of 2-parametric family of
multifractal density invariants.  In fact, it is easy to prove the
following theorem:  If there exist a and b such that

\begin{eqnarray}\frac{< \rho_\epsilon{}^{aq} (x) >}{< \rho_\epsilon{}^a
(x)>^q < \rho_\epsilon {}^q (x) >^b} = const \end{eqnarray}

is independent on scale  at $q \in R $, then $\rho_\epsilon (x)$ is a
multifractal density of universal multifractal with the Levy index
$\alpha =\frac{\ln b}{\ln a}$. The proof of this theorem can easily be
obtained if one consider Eq. (14) as a functional equation for the
momentum scaling function and repeat the solution procedure this equation
in a way analogous to that sited above.  Inverse theorem is also valid,
its proof being trivial after substitution of $K (q)$ for universal
multifractals.  This, the proposed scheme of multifractal density
 interpolation is based on the existence of the invariants (14)
 (independent of scale resolution) for universal multifractal
 distributions. Such invariants can be useful for the study of
 multifractal properties of the turbulence and of other strongly non
 equilibrium systems, because these invariants establish relation between
  the moments of different orders and determine splitting the moments for
  universal multifractals.

In conclusion we discuss convergence of the proposed algorithm. For this
purpose it is convenient to rewrite Eq.(9) for the scaling function. The
rewritten equation describes the scaling function evolution during the
iteration process,
		  \begin{eqnarray}K_{n+1} (q) = \frac{1}{2} ( K_n (q\nu )
		  - q K_n (\nu ) ) \end{eqnarray}
where $K_{n+1} (q)$ is the scaling function at the $(n + 1)$-th iteration
      step. With the use of Eq.(15) we consider the scaling function at
      the $n$-th iteration step with the initial condition taken into
      account.  \begin{eqnarray}K_1 (q)= \frac{K_0}{\alpha - 1} (q^\alpha
      - q) + \Delta (q) \end{eqnarray} The function $\Delta (q)$ is caused
 by initial deviation from the scaling function of the universal
multifractal. Then, after $n$ iterations we get \begin{eqnarray}K_n (q) =
\frac{K_0}{\alpha -1} (\frac{\nu^{\alpha n}}{2^n}) (q^\alpha - q ) +
\frac{1}{2^n} (\Delta (q\nu^n ) - q \Delta (\nu^n ) ) \end{eqnarray} It
follows from Eq.(17) that for $\nu^\alpha =2 $ \begin{eqnarray}K_n (q)
\stackrel{ n\to \infty }{\longrightarrow} \frac{K_0}{\alpha -1} (q^\alpha
-q)  \end{eqnarray} if $\Delta (q)$ is bounded above and below (i.e. $C_1
< \Delta (q) < C_2 $ ). Moreover, if $\Delta (q) $ grows not faster than
const$ q^\beta $ with $q$ growing, where $\beta < \alpha$, and $\beta =
1$, is also included, then the scaling function also converges to the
scaling function of universal multifractal.  Thus, it is proved that the
moments of the interpolated multifractal converge to those of the
universal multifractal (of course, if initial deviations do not grow
rather fast with q drawing). It implies that the proposed algorithm can be
used for interpolation of the universal multifractals. It is noteworthy
that the simplicity of this algorithm also allows one to the use it
effectively for studying properties of the universal multifractals by
purely analytical methods.  Therefore, the proposed algorithm allows one
to interpolate experimental data and reconstruct multifractal
interpolating functions, which belong to universal multifractals, with
arbitrary index $\alpha$. It allows one to use the algorithm for the
description and creation of the models of various natural processes and
structures, in particular, ecologically important distributions of
technogenic admixtures precipitated after the transport in turbulent
atmosphere.  \vspace{5mm}

{\bf Results }\\
      One - parameter family of invariants of universal multifractals has
      been obtained. The existence of such invariants implies the existence
      of rules allowing one to express higher one-point moments in terms
      of lower ones. \\
      If for a multifractal one finds one of these invariants,this
      signifies that such a multifractal is in class of the universal
      multifractals with a definite Levy index. A simple algorithm for
      interpolating universal multifractals based on the invariance
      discovered has been proposed. The convergence of the algorithm has
      been proved, and velocity of convergence has been analytically
      estimated. \\

 {\bf Acknowledgements }\\
 This work was supported by
     	   International Association under the project INTAS 93-1194.

\end{document}